\documentclass[aps,prc,preprint,showpacs]{revtex4-1}  
\usepackage{graphicx}
\usepackage{amsmath}
\usepackage{color}
\begin{document}
\newcommand {\nc} {\newcommand}
\nc {\beq} {\begin{eqnarray}} 
\nc {\eol} {\nonumber \\} 
\nc {\eeq} {\end{eqnarray}} 
\nc {\eeqn} [1] {\label{#1} \end{eqnarray}}   
\nc {\eoln} [1] {\label{#1} \\} 
\nc {\ve} [1] {\mbox{\boldmath $#1$}}
\nc {\rref} [1] {(\ref{#1})} 
\nc {\Eq} [1] {Eq.~(\ref{#1})} 
\nc {\re} [1] {Ref.~\cite{#1}} 
\nc {\la} {\langle} 
\nc {\ra} {\rangle} 
\nc {\cA} {\mathcal{A}} 
\nc {\cM} {\mathcal{M}} 
\nc {\cY} {\mathcal{Y}} 
\nc {\dem} {\mbox{$\frac{1}{2}$}} 
\nc {\ut} {\mbox{$\frac{1}{3}$}} 
\nc {\qt} {\mbox{$\frac{4}{3}$}} 
\nc {\arrow} [2] {\mbox{$\mathop{\rightarrow}\limits_{#1 \rightarrow#2}$}}

\author{D. Baye}
\email[]{dbaye@ulb.ac.be}
\affiliation{Physique Quantique, and Physique Nucl\'eaire Th\'eorique et Physique Math\'ematique, \\ 
C.P. 229, Universit\'e libre de Bruxelles (ULB), B-1050 Brussels Belgium.}
\author {E.M. Tursunov}
\email[]{tursune@inp.uz}
\affiliation {Institute of Nuclear Physics, Uzbekistan Academy of Sciences, \\
100214, Ulugbek, Tashkent, Uzbekistan}

\title{Isospin-forbidden electric-dipole capture and the $\alpha(d,\gamma)^6$Li reaction}

\date{\today}

\begin{abstract}
At the long-wavelength approximation, $E1$ transitions are forbidden between isospin-zero states. 
Hence $E1$ radiative capture is strongly hindered in reactions involving $N = Z$ nuclei 
but the $E1$ astrophysical $S$ factor may remain comparable to, or larger than, the $E2$ one. 
Theoretical expressions of the isoscalar and isovector contributions to $E1$ capture are analyzed 
in microscopic and three-body approaches in the context of the $\alpha(d,\gamma)^6$Li reaction. 
The lowest non-vanishing terms of the operators are derived 
and the dominant contributions to matrix elements are discussed. 
The astrophysical $S$ factor computed with some of these contributions in a three-body $\alpha+n+p$ model 
is in agreement with the recent low-energy experimental data of the LUNA collaboration. 
This confirms that a correct treatment of the isovector $E1$ transitions 
involving small isospin-one admixtures in the wave functions should be able 
to provide an explanation of the data without adjustable parameter. 
The exact-masses prescription which is often used to avoid the disappearance 
of the $E1$ matrix element in potential models is not founded at the microscopic level 
and should not be used for such reactions. 
The importance of capture components from an initial $S$ scattering wave is also discussed. 
\end{abstract}

\maketitle

\section{Introduction}

In some radiative-capture reactions between light nuclei,
electric-dipole transitions are strongly suppressed \cite{AAR99}. 
This effect is due to an isospin selection rule: 
$E1$ transitions are isospin-forbidden in capture reactions involving $N = Z$ nuclei \cite{Te67}. 

At the long-wavelength approximation, which is a good approximation for this type of reactions, 
the isoscalar part of the $E1$ operator vanishes and transitions take place via its isovector part. 
Matrix elements of isovector operators vanish between isospin-zero states. 
However, except for the deuteron, 
realistic wave functions of $N = Z$ nuclei are not pure eigenstates of the isospin operator 
and $E1$ transitions are not exactly forbidden. 
Their strength may keep an order of magnitude 
similar to the strength of the usually much weaker $E2$ transitions. 
This effect is particularly spectacular for the $^{12}$C($\alpha,\gamma)^{16}$O reaction 
where the isospin-forbidden $E1$ component is enhanced by resonances (see references in Ref.~\cite{AAR99}). 
Disentangling the $E1$ and $E2$ strengths is experimentally very difficult 
and the theoretical calculations of the $E1$ component are still quite uncertain. 
The role of $E1$ transitions is also complicated in other reactions of astrophysical interest such as 
$d(d,\gamma)^4$He, $^4$He$(d,\gamma)^6$Li, and $^{16}$O$(\alpha,\gamma)^{20}$Ne. 
It may also play some role in the triple $\alpha$ mechanism generating $^{12}$C. 

An {\it ab initio} description of the two lightest cases is in principle possible at present. 
The astrophysical $S$ factor of the $d(d,\gamma)^4$He reaction has been computed 
with an {\it ab initio} calculation in Ref.~\cite{AAS11}. 
The $E1$ component is mainly obtained from $T = 1$ isospin components in $^4$He introduced 
by coupled $p + ^3$H and $n + ^3$He configurations. 
Its largest contribution reaches at most 4\% near the center-of-mass energy 0.01 MeV 
and thus remains quite small with respect to $E2$ \cite{DBS14}. 
For the $^4$He$(d,\gamma)^6$Li reaction, the problem is more difficult 
because of the larger numbers of nucleons and of possible configurations. 
An {\it ab initio} study of the $\alpha + d$ elastic scattering has been performed 
in Ref.~\cite{NQ11} with a realistic nucleon-nucleon ($NN$) force. 
A study of the $E2$ capture component could be based on that work but the study of the $E1$ component 
would require much additional computer time with the introduction of $T = 1$ isospin components 
in the initial and final wave functions. 
Such a calculation is thus not available yet. 

Preliminary attempts to calculate isospin-forbidden $E1$ cross sections for heavier systems  
have been performed in microscopic cluster models. 
In Ref.~\cite{DB86}, an $\alpha$ cluster with a small $T = 1$ component in its ground-state  
has been used to explore $E1$ capture in the $^{16}$O$(\alpha,\gamma)^{20}$Ne reaction 
but a similar component would at least have been necessary in the $^{16}$O cluster. 
In Ref.~\cite{DB87}, $E1$ capture in the $^{12}$C($\alpha,\gamma)^{16}$O reaction was studied 
by coupling $^{12}$C$ + \alpha$ configurations with $^{15}$N$+p$ and $^{15}$O$+n$ 
configurations which introduced some $T = 1$ contributions in the 16-nucleon wave functions 
but some properties of the $E1$ resonances had to be modified phenomenologically. 
These attempts provide qualitative information but remain too limited for quantitative predictions. 

Since realistic microscopic calculations are not available yet, 
most calculations of isospin-forbidden $E1$ capture have been performed 
in the two-body or potential model based on the cluster idea. 
The isospin quantum number does not appear in this model. 
The nuclei are only represented by their atomic numbers $Z_1$ and $Z_2$, 
their mass numbers $A_1$ and $A_2$, and their spin and parity quantum numbers. 
The physics arises from the interaction between them. 
Electric dipole transitions are nevertheless forbidden because of the presence 
of a factor $Z_1/A_1 -Z_2/A_2$ in $E1$ transition matrix elements, 
which vanishes for $N = Z$ colliding nuclei since both ratios $Z_1/A_1$ and $Z_2/A_2$ are equal to 1/2. 
Indeed, this factor in the effective $E1$ operator is of microscopic origin 
and thus involves \textit{integer} mass numbers. 

In order to have a non-vanishing $E1$ astrophysical $S$ factor, the traditional prescription is 
to replace the integer mass numbers $A_1$ and $A_2$ by non-integer values 
deduced from the experimental masses of the colliding nuclei. 
This replacement is usually justified by the fact 
that it leads to a non-vanishing dipole moment of the nucleus in the cluster picture. 
This `exact-masses' prescription, however, has no microscopic foundation at the nucleon level. 
As discussed below, it may give a plausible order of magnitude for the capture cross section 
but the possible agreement or disagreement with experimental data has no physical meaning. 
The energy dependence of the cross section may also be plausible but is not founded microscopically. 

In this paper, we discuss various theoretical aspects of the forbidden $E1$ transitions. 
To fix ideas, we take the $\alpha + d \rightarrow ^6$Li$ + \gamma$ capture process as an example. 
This reaction was first studied experimentally at energies around and above the 0.711 MeV $3^+$ resonance 
\cite{robe81,mohr94}. 
Until recently, the lower-energy data resulted from indirect measurements 
with Coulomb breakup reactions of $^6$Li on lead \cite{kien91,hamm10}. 
The presence of nuclear breakup makes difficult the extraction of information 
on radiative capture from the data. 
Recently, the $\alpha(d,\gamma)^6$Li reaction was studied at the LUNA facility 
by direct measurements at the two astrophysical energies 94 and 134 keV \cite{luna14}. 

From the theoretical side, calculations of astrophysical $S$ factors have been developed 
within different two-body potential models
\cite{LR86,BZZ90,Ja93,MST95,dub95,mukh11,tur15,MSB16,GMM17}, 
three-body potential models \cite{CET90,RES95,TKT16}, 
and with semi-microscopic \cite{lang86,noll01} 
and microscopic \cite{TBL91,desc98} models. 
Early models focused on the then existing data \cite{robe81} 
at energies around and beyond the $3^+$ resonance 
where the main contribution to the capture process comes from $E2$ transitions. 
At low energies, the dominant contribution is expected to come from the $E1$ transition operator 
since the $E2$ cross section is smaller than the data in all models. 
The recent LUNA data have renewed the interest for theoretical calculations of the $S$ factor 
at astrophysical energies \cite{MSB16,TKT16,GMM17}. 

In the theoretical literature, the $E1$ capture is treated in various ways, 
but the exact-masses prescription is in general used in potential models 
\cite{BZZ90,RES95,MST95,dub95,mukh11,tur15,MSB16,TKT16,GMM17} 
and even in partly microscopic approaches \cite{lang86,TBL91,noll01}, 
sometimes combined with various other corrections. 
These calculations raise questions about the foundation of the exact-masses prescription 
and about the validity of its combination with other corrections. 

The aim of the present study is to discuss theoretical aspects of the forbidden $E1$ transitions 
and question the validity of the exact-masses prescription. 
We analyze theoretically different contributions to the $E1$ $S$ factor 
of the $\alpha(d,\gamma)^6$Li capture process 
and emphasize the main ones that should be necessarily included in a realistic model. 
A model able to take all these contributions into account in a consistent way is beyond our reach. 
We evaluate some of these contributions to the $S$ factor 
with the three-body $\alpha+n+p$ model of Ref.~\cite{TKT16} to discuss their importance. 
This allows us to suggest key points that should be studied in future model calculations. 

In Sec.~\ref{sec:edift}, the microscopic expression of the electric dipole operator 
and the corresponding matrix elements for isospin-forbidden transitions are presented. 
In Sec.~\ref{sec:tbmift}, the expressions are specialized to a three-body model. 
The initial wave function is the product of a two-body deuteron wave function 
and an $\alpha+d$ scattering wave function. 
The final $^6$Li$(1^+)$ ground state is described with an $\alpha+n+p$ three-body wave function 
in hyperspherical coordinates \cite{DDB03,tur06}. 
The model involves $n + p$, $\alpha + n/p$, and $\alpha + d$ potentials. 
In Sec.~\ref{sec:nr}, results are presented and commented. 
The exact-masses prescription is discussed in Sec.~\ref{sec:emp} 
as well the possible role of capture from an initial $S$ wave. 
Sec.~\ref{sec:conc} is devoted to a conclusion. 

\section{Microscopic treatment of isospin-forbidden $E1$ transitions}
\label{sec:edift}
\subsection{Microscopic electric multipole operators}
\label{sec:mo}
Since the energies of the emitted photons are usually not large at astrophysical energies, 
their wavelengths are large with respect to typical dimensions of the system 
and the photon wavenumbers 
\beq
k_\gamma = E_\gamma/\hbar c
\eeqn{2.2}
can be considered as small. 
The long-wavelength approximation can be used. 
Let $\ve{r}_j$ be the coordinate of the $j$th nucleon. 
At the long-wavelength approximation, the translation-invariant 
electric transition operators of multipolarity $\lambda$ are given to a good approximation by  
\beq
{\cal M}_\mu^{E\lambda} = e \sum_{j=1}^A (\dem - t_{j3}) r^{\prime\lambda}_j Y_{\lambda\mu} (\Omega'_j),
\eeqn{2.a}
where $t_{j3}$ is the third component of the isospin operator $\ve{t}_j$ of the $j$th nucleon 
related to its charge by $e(\dem-t_{j3})$, and 
\beq
\ve{r}'_j = \ve{r}_j - \ve{R}_{\rm cm}
\eeqn{2.b}
is its coordinate with respect to the center of mass 
\beq
\ve{R}_{\rm cm} = \frac{1}{A} \sum_{j=1}^A \ve{r}_j
\eeqn{2.c}
of the $A$-nucleon system. 
The functions $Y_{\lambda\mu} (\Omega'_j)$ are spherical harmonics depending 
on the angular part of $\ve{r}'_j = (r'_j,\Omega'_j)$. 

The orbital angular momentum with respect to the center of mass and spin of nucleon $j$ 
are denoted as $\ve{L}'_j$ and $\ve{S}_j$, respectively. 
The total orbital momentum operator of the system is $\ve{L} = \sum_{j=1}^A \ve{L}'_j$, 
the total spin is $\ve{S} = \sum_{j=1}^A \ve{S}_j$ 
and the total angular momentum is $\ve{J} = \ve{L} + \ve{S}$. 
The total isospin operator of the system is $\ve{T} = \sum_{j=1}^A \ve{t}_j$. 

The operators defined by \Eq{2.a} contain isoscalar (IS) and isovector (IV) parts. 
At the long-wavelength approximation, the $E1$ operator is special. 
It mainly contains an isovector component, 
\beq
{\cal M}_\mu^{E1} \approx {\cal M}_\mu^{E1,\textrm{IV}} = -e \sum_{j=1}^A t_{j3} r'_j Y_{1\mu} (\Omega'_j).
\eeqn{2.1}
The lowest-order term of the isoscalar part vanishes since $\sum_{j=1}^A \ve{r}'_j = 0$. 
This operator connects eigenstates of the total isospin operator 
with initial and final isospin quantum numbers differing by one unit, $T_f = |T_i \pm 1|$. 
It also connects states with $T_i = T_f$, but only for $N \neq Z$. 
Transitions from $T_i = 0$ to $T_f = 0$ are forbidden. 

The isoscalar part of the $E1$ operator is however not exactly zero. 
It might play a non-negligible role in some cases. 
The first non-vanishing term reads using the Siegert theorem \cite{Ba12} 
\beq
{\cal M}_\mu^{E1,\textrm{IS}} & \approx & 
- \frac{1}{60} e k_\gamma^2 \sum_{j=1}^A r_j^{\prime 3} Y_{1\mu} (\Omega'_j)
\eol
&& + \frac{e\hbar k_\gamma}{8 m_p c} \sum_{j=1}^A r'_j [\ve{L} Y_{1\mu}] (\Omega'_j) \cdot 
\left[\frac{2}{3} \ve{L}'_j + (g_p+g_n) \ve{S}_j \right].
\eeqn{2.3}
where $m_p$ is the proton mass, and $g_p$ and $g_n$ are the proton and neutron 
gyromagnetic factors, respectively. 
The vector function $[\ve{L} Y_{1\mu}] (\Omega)$ is the result of the action of the orbital momentum operator 
on the spherical harmonics $Y_{1\mu} (\Omega)$ with $l = 1$. 
This operator connects components with the same initial and final isospins, $T_i = T_f$. 
When it acts on a wave function with a largely dominant component 
with zero total orbital momentum and small intrinsic spin,
the first term of \Eq{2.3} should give a reasonable approximation. 
\subsection{Transition matrix elements}
\label{sec:tme}
We consider transitions in $N = Z$ systems between an initial scattering state 
and a final bound state with dominant zero-isospin components. 
Their wave functions can be written symbolically as 
\beq
\Psi^{JM}_{i,f} = \Psi^{JM;0}_{i,f} + \Psi^{JM;1}_{i,f}.
\eeqn{2.10}
The $T = 1$ components $\Psi^{JM;1}_{i,f}$ are much smaller than the $T = 0$ components 
$\Psi^{JM;0}_{i,f}$. 
Possible admixtures of larger isospin values are neglected. 

To a good approximation, three types of matrix elements must be calculated. 
Two of them involve an isovector transition, 
\textit{i.e.}, between the dominant $T_i = 0$ component in the initial scattering state 
and the $T_f = 1$ admixture in the final bound state  
\beq
\la \Psi^{J'M';1}_{f} | {\cal M}_\mu^{E1,\textrm{IV}} | \Psi^{JM;0}_{i} \ra,
\eeqn{2.11}
and between the $T_i = 1$ admixture in the initial scattering state 
and the dominant $T_f = 0$ component in the final bound state  
\beq
\la \Psi^{J'M';0}_{f} | {\cal M}_\mu^{E1,\textrm{IV}} | \Psi^{JM;1}_{i} \ra.
\eeqn{2.12}
An isoscalar transition is also possible, essentially between the dominant components, 
\beq
\la \Psi^{J'M';0}_{f} | {\cal M}_\mu^{E1,\textrm{IS}} | \Psi^{JM;0}_{i} \ra.
\eeqn{2.13}
The $E1$ transition matrix element is the coherent sum of these three contributions. 
\subsection{$\alpha(d,\gamma)^6$Li $E1$ capture in resonating-group notation}
\label{sec:adrgm}
To fix ideas we consider the $\alpha(d,\gamma)^6$Li reaction. 
We use the notation of the resonating-group method (RGM) \cite{WT77,Ta81}. 
This notation is also valid for {\it ab initio} descriptions. 
We limit ourselves to $\alpha + n + p$ configurations. 
Realistic calculations might also include $^3$H$ + ^3$He configurations, for example, 
that we neglect to simplify the presentation.  
The wave functions that we now describe display the main components expected to play 
a significant role in $E1$ transitions. 
Many other smaller components are of course possible. 

In the RGM, a partial wave of the initial scattering wave function \rref{2.10} is written as 
\beq
\Psi^{JM\pi}_{i} = \cA \phi^{00+}_\alpha [\phi^{1+}_d \otimes Y_L(\Omega_R)]^{JM} g_i^{J\pi}(R),
\eeqn{2.20}
where $\cA$ is the six-nucleon antisymmetrizer and $\ve{R} = (R,\Omega_R)$ is the relative 
coordinate between the centers of mass of the $\alpha$ and deuteron clusters. 
The functions $\phi^{00+}_\alpha$ and $\phi^{1m+}_d$ 
are translation-invariant internal wave functions of the ground states 
of the $^4$He nucleus with angular momentum 0 and positive parity 
and of the deuteron with angular momentum 1 and positive parity, respectively. 
The $^4$He wave function depends on three internal coordinates. 
The deuteron wave function depends on the relative coordinate $\ve{r}= (r,\Omega_r)$ 
between the proton and neutron. 
The total parity $\pi$ is equal to $(-1)^L$.  
The $^4$He ground-state internal wave function may contain a small $T = 1$ admixture 
\beq
\phi^{00+}_\alpha = \phi^{00+;0}_\alpha + \phi^{00+;1}_\alpha.
\eeqn{2.21}
The $T = 1$ component is mainly due to the Coulomb interaction between the protons. 
The neutron-proton mass difference and isospin non-conserving terms in the nuclear 
force also contribute but to a lesser extent. 
The deuteron ground-state wave function is purely $T = 0$. 
In reactions of $\alpha$ particles with heavier $N = Z$ nuclei, 
a $T = 1$ admixture also appears in the second cluster. 

Various corrections may also appear in the scattering wave function  
to take distortion of the initial state at short distances into account. 
They may involve sums over pseudo-states of the deuteron and/or of the $\alpha$ particle. 
The most important ones should arise from deuteron pseudo-states 
which can simulate its Coulomb polarizability \cite{Ja93}.  
They may also include additional shell-model-like $^6$Li terms \cite{WT77}.  
We do not display these corrections here to simplify the discussion 
but they can be treated in the same way as similar terms displayed below in the final state. 

Under some simplifying assumptions, 
the main components of the final bound-state wave function of the $1^+$ ground state of $^6$Li 
can be approximated as  
\beq
\Psi^{1M+}_{f} & = & \cA \phi^{00+}_\alpha [\phi^{1+}_d \otimes Y_0(\Omega_R)]^{1M} g_f^{1+}(R) 
\eol
&& + \sum_n \cA \phi^{00+}_\alpha [\phi^{1\pi_n;T_n}_{d^*n} \otimes Y_{L_n}(\Omega_R)]^{1M} g^{1+}_{d^*n}(R) 
\eol
&& + \sum_{I,n} \cA [[\phi^{1-;1}_{\alpha^*n} \otimes \phi^{1+}_{d}]^I 
\otimes Y_1(\Omega_R)]^{1M} g^{1+}_{\alpha^* In}(R).
\eeqn{2.23}
The $\phi^{1\pi_n;T_n}_{d^*n}$ with $T_n = 0$ or 1 are excited pseudo-states of the deuteron.  
The relative orbital momentum is $L_n = 0$ for $\pi_n = +$ and $L_n = 1$ for $\pi_n = -$. 
The $\phi^{1-;1}_{\alpha^*n}$ are excited pseudo-states of the $^4$He nucleus 
with angular momentum 1 and isospin 1. 
The channel spin $I$ can take the values 0, 1, and 2. 

Given the angular momentum and parity of the final state, 
the initial state for $E1$ transitions corresponds to $J = 0$, 1 and 2 and a negative parity. 
This is realized by choosing $L = 1$ in \Eq{2.20}. 
Within these assumptions, let us write the various matrix elements. 
Matrix element \rref{2.11} reads for an initial wave with $L = 1$,
\beq
&& \la \Psi^{1M'+;1}_{f} | {\cal M}_\mu^{E1,\textrm{IV}} | \Psi^{JM-;0}_{i} \ra 
\eol
&& = \la \cA \phi^{00+;1}_\alpha [\phi^{1+}_d \otimes Y_0]^{1M'} g_f^{1+}(R) 
| {\cal M}_\mu^{E1,\textrm{IV}} | 
\cA \phi^{00+;0}_\alpha [\phi^{1+}_d \otimes Y_1]^{JM} g_i^{J-}(R) \ra 
\eol
&& + \sum_n \cA \phi^{00+;0}_\alpha [\phi^{1\pi_n;T_n}_{d^*n} \otimes Y_{L_n}]^{1M'} g^{1+}_{d^*n}(R) 
| {\cal M}_\mu^{E1,\textrm{IV}} | 
\cA \phi^{00+;0}_\alpha [\phi^{1+}_d \otimes Y_1]^{JM} g_i^{J-}(R) \ra 
\eol
&& + \sum_{I,n} \cA [[\phi^{1-;1}_{\alpha^*n} \otimes \phi^{1+}_{d}]^I 
\otimes Y_1]^{1M'} g^{1+}_{\alpha^* In}(R) 
| {\cal M}_\mu^{E1,\textrm{IV}} | 
\cA \phi^{00+;0}_\alpha [\phi^{1+}_d \otimes Y_1]^{JM} g_i^{J-}(R) \ra
\eeqn{2.31}
and matrix element \rref{2.12} reads 
\beq
&& \la \Psi^{1M'+;0}_{f} | {\cal M}_\mu^{E1,\textrm{IV}} | \Psi^{JM-;1}_{i} \ra 
\eol
&& = \la \cA \phi^{00+;0}_\alpha [\phi^{1+}_d \otimes Y_0]^{1M'} g_f^{1+}(R)
| {\cal M}_\mu^{E1,\textrm{IV}} 
| \cA \phi^{00+;1}_\alpha [\phi^{1+}_d \otimes Y_1]^{JM} g_i^{J-}(R) \ra,
\eeqn{2.32}
where $J$ can be equal to 0, 1 and 2. 
Other contributions appear when the initial state is distorted. 
Matrix element \rref{2.13} reads 
\beq
&& \la \Psi^{1M'+;0}_{f} | {\cal M}_\mu^{E1,\textrm{IS}} | \Psi^{JM-;0}_{i} \ra 
\eol
&& = \la \cA \phi^{00+;0}_\alpha [\phi^{1+}_d \otimes Y_0]^{1M'} g_f^{1+}(R) 
| {\cal M}_\mu^{E1,\textrm{IS}} | 
\cA \phi^{00+;0}_\alpha [\phi^{1+}_d \otimes Y_1]^{JM} g^{J\pi}(R) \ra. 
\eeqn{2.33}
As the operator is much smaller here, only the dominant $T = 0$ components are kept. 
\section{Three-body model of isospin-forbidden $E1$ transitions}
\label{sec:tbmift}
\subsection{Three-body $E\lambda$ operators}
\label{sec:tbo}
We now consider the three-body $\alpha + n + p$ model. 
The $^4$He nucleus is treated as a structureless particle. 
Its properties appear in the interaction with the nucleons. 
They may also appear in some parameters of the model. 

Let us start from the isovector microscopic operator \rref{2.1}. 
Let us assume that the first four coordinates $\ve{r}_j$ correspond to the $\alpha$ particle 
and that the last two correspond to the deuteron. 
In vector notation, operator \rref{2.1} reads 
\beq
\ve{\cal M}^{E1,\textrm{IV}} = -e \sum_{j=1}^6 t_{j3} (\ve{r}_j - \ve{R}_{\rm cm}).
\eeqn{3.1}
The deuteron internal coordinate is 
\beq
\ve{r} = \ve{r}_5 - \ve{r}_6
\eeqn{3.2}
and the $\alpha$-deuteron relative coordinate is given by 
\beq
\ve{R} = \ve{R}^\alpha_{\rm cm} - \dem (\ve{r}_5 + \ve{r}_6),
\eeqn{3.4}
where $\ve{R}^\alpha_{\rm cm} = \frac{1}{4} \sum_{j=1}^4 \ve{r}_j$ 
is the center-of-mass coordinate of the $\alpha$ particle.

Then, the $E1$ operator can be rewritten as 
\beq
\ve{\cal M}^{E1,\textrm{IV}} = \ve{\cal M}_\alpha^{E1,\textrm{IV}} 
- \dem\, e (t_{5,3} - t_{6,3}) \ve{r} 
- \ut\, e (T_{\alpha 3} - 2 T_{d 3}) \ve{R},
\eeqn{3.10}
where the first term 
\beq
\ve{\cal M}_\alpha^{E1,\textrm{IV}} = -e \sum_{j=1}^4 t_{j3} (\ve{r}_j - \ve{R}^\alpha_{\rm cm})
\eeqn{3.11}
is the $E1$ operator for the $\alpha$ particle. 
The second term is the $E1$ operator for the deuteron 
and the last term corresponds to the relative motion. 
The operators $\ve{T}_\alpha = \sum_{j=1}^4 \ve{t}_j$ and $\ve{T}_d = \ve{t}_5 + \ve{t}_6$ 
are the isospin operators of the $\alpha$ particle and deuteron, respectively. 
Hence, in multipolar form, one has 
\beq
\cM_\mu^{E1,\textrm{IV}} = \cM_{\alpha,\mu}^{E1,\textrm{IV}} 
- \dem\, e (t_{5,3} - t_{6,3}) \cY_{1\mu}(\ve{r}) 
- \ut\,  e (T_{\alpha 3} - 2 T_{d 3}) \cY_{1\mu}(\ve{R}) 
\eeqn{A1}
with
\beq
\cY_{\lambda\mu}(\ve{x}) = x^\lambda Y_{\lambda\mu}(\Omega_x).
\eeqn{A8}
For more general clusters with mass numbers $A_1$ and $A_2$, 
the factor in front of $-e \cY_{1\mu}(\ve{R})$ in the last term becomes 
$(A_2 T_{A_1 3} - A_1 T_{A_2 3})/A$. 
Its eigenvalue contains the factor $Z_1/A_1 - Z_2/A_2$ mentioned in the introduction. 

In a similar way, the first term of the isoscalar $E1$ operator \rref{2.3} becomes 
\beq
\cM_\mu^{E1,\textrm{IS}} = \cM_{\alpha,\mu}^{E1,\textrm{IS}} - \frac{1}{60} k_\gamma^2\, 
\left\{ \frac{5}{9} e (4R_\alpha^2-R^2-r^2) \cY_{1\mu}(\ve{R}) \qquad \qquad \right. 
\eol 
\left. - \frac{\sqrt{32\pi}}{9} \left( 2 [\cY_1(\ve{R}) \otimes \cM_\alpha^{E2,\textrm{IS}}]_{1\mu} 
- e [\cY_1(\ve{R}) \otimes \cY_2(\ve{r})]_{1\mu} \right) \right\},
\eeqn{A2}
where $R_\alpha^2 = \frac{1}{4} \sum_{j=1}^4 (\ve{r}_j - \ve{R}^\alpha_{\rm cm})^2$, 
and the $E2$ operator reads 
\beq
\cM_\mu^{E2} = \cM_{\alpha,\mu}^{E2} 
+ \frac{\sqrt{120\pi}}{9} [\cY_1(\ve{R}) \otimes \cM_\alpha^{E1,\textrm{IV}}]_{2\mu} 
+ \frac{1}{9} e (6 - T_{\alpha 3} - 4T_{d 3}) \cY_{2\mu}(\ve{R}) 
\eol
+ \frac{1}{4} e (1 - T_{d 3}) \cY_{2\mu}(\ve{r}) 
+ \frac{\sqrt{120\pi}}{9} e (t_{5,3} - t_{6,3}) [\cY_1(\ve{R}) \otimes \cY_1(\ve{r})]_{2\mu}.
\eeqn{A3}

In the simplest version of a three-body model, the $\alpha$ particle is in its ground state $\phi_\alpha^{00+}$. 
Effective multipole operators are obtained by taking the mean value of the above expressions, 
\beq
\widetilde{\cM}_\mu^{E\lambda} = \la \phi_\alpha^{00+} | \cM_\mu^{E\lambda} | \phi_\alpha^{00+} \ra.
\eeqn{A4}
The eigenvalue of $T_{\alpha 3}$ is zero, as well as the mean value of $\cM_{\alpha,\mu}^{E\lambda}$. 
The eigenvalue of $T_{d 3}$ vanishes for the neutron-proton system. 
Hence, for $E1$, one obtains from \rref{A1} and \rref{A2}, 
with the neutron as particle 5 and the proton as particle 6, 
\beq
\widetilde{\cM}_\mu^{E1,\textrm{IV}} = \frac{1}{2} e \cY_{1\mu}(\ve{r}) 
\eeqn{A5}
and 
\beq
\widetilde{\cM}_\mu^{E1,\textrm{IS}} = - \frac{1}{60} e k_\gamma^2 \left\{
 \frac{5}{9} (4r_\alpha^2-R^2-r^2) \cY_{1\mu}(\ve{R}) 
+ \frac{\sqrt{32\pi}}{9} [\cY_1(\ve{R}) \otimes \cY_2(\ve{r})]_{1\mu} \right\},
\eeqn{A6}
where $r_\alpha^2 = \la \phi_\alpha^{00+} | R_\alpha^2 | \phi_\alpha^{00+} \ra$ 
is the mean square radius of the $\alpha$ particle. 
With \rref{A3}, the $E2$ operator is given by 
\beq
\widetilde{\cM}_\mu^{E2} = \frac{2}{3} e \cY_{2\mu}(\ve{R}) + \frac{1}{4} e \cY_{2\mu}(\ve{r}) 
- \frac{\sqrt{120\pi}}{9} e [\cY_1(\ve{R}) \otimes \cY_1(\ve{r})]_{2\mu}.
\eeqn{A7}
This expression can also be deduced from Eq.~(B2) of \re{DDB03}. 
The first two terms are also derived in \re{CET90}. 
\subsection{Transition matrix elements}
\label{sec:tbtme}
In the present $\alpha + n + p$ three-body model, the initial scattering wave function is defined 
by coupling the ground-state deuteron wave function with partial waves describing the relative motion. 
The polarizability of the deuteron and other distortion effects of the initial wave are thus neglected. 
The deuteron wave function is defined as a pure $s$ state (except in Sec.~\ref{sec:S} below) by 
\beq
\phi^{lSjm}(\ve{r}) = [Y_l(\Omega_r) \otimes \chi^S]^{jm} r^{-1} u^{lSj}(r)
\eeqn{3.22}
with $l = 0$ and $S = j = 1$. 
The spinor $\chi^S$ is the total spin state of the neutron and proton. 
The initial scattering functions for partial wave $L$ read 
\beq
\Psi^{JM\pi}_i(\ve{r},\ve{R}) = [\phi^{011+}_d(\ve{r}) \otimes \Phi^{L\pi}(\ve{R})]^{JM}
\eeqn{3.21}
with $\pi = (-1)^L$ and 
\beq
\Phi^{Lm\pi}(\ve{R}) = Y_{Lm}(\Omega_R) g_i^{L\pi}(R),
\eeqn{3.23}
since the $\alpha$ particle has spin 0 and positive parity. 

The final $^6$Li$(1^+)$ ground state is described by a three-body wave function 
defined in the hyperspherical basis as 
\beq
\Psi_{f}^{1M+}(\ve{r},\ve{R}) = 
\rho^{-5/2} \sum_{\gamma, K} \chi_{\gamma K}(\rho) {\cal Y}^{1M}_{\gamma K}(\Omega_5)
\eeqn{4.1}
where $\rho = \sqrt{\dem r^2 + \qt R^2}$ is the hyperradius and $\Omega_5$ represents five angles, 
the orientations $\Omega_r$ of $\ve{r}$ and $\Omega_R$ of $\ve{R}$, 
and the hyperangle $\alpha = \arctan (\sqrt{8/3}\, R/r)$ (see Refs.~\cite{DDB03,tur06} for details). 
Number $K$ is the hypermomentum. 
Notation $\gamma$ represents the other quantum numbers of the problem, 
\textit{i.e.}, the orbital momentum $l$ and spin $S$ of the proton-neutron pair, 
and the orbital momentum $L$ of the $\alpha-(n+p)$ relative motion. 
The functions ${\cal Y}^{JM}_{\gamma K}(\Omega_5)$ are hyperspherical harmonics 
and the functions $\chi_{\gamma K}(\rho)$ are hyperradial functions. 
The positive parity requires $l + L$ even. 

Thanks to the antisymmetry of the deuteron wave function, 
it is possible to associate an isospin to the different parts of the three body wave function, 
\beq
\Psi^{1M+}_f(\ve{r},\ve{R}) = \Psi^{1M+;0}_f(\ve{r},\ve{R}) + \Psi^{1M+;1}_f(\ve{r},\ve{R}).
\eeqn{3.24}
For the neutron-proton system in the isospin formalism, antisymmetry imposes that $l + S + T$ must be odd. 
Hence it is possible to perform the separation \rref{3.24} 
of the final wave function according to the deuteron isospin $T$. 
The component with $l + S$ odd corresponds to $T_f = 0$ 
while the component with $l + S$ even corresponds to $T_f = 1$. 
The wave function \rref{4.1} can be interpreted as corresponding to the first two terms of \Eq{2.23}. 
Indeed, while the $\alpha$ particle is frozen in its ground state, 
the deuteron can be fully distorted or excited 
and $T_f = 1$ admixtures can appear in the neutron-proton system. 

Matrix element \rref{2.31} becomes with \rref{3.10}, 
\beq
\la \Psi^{1M'+;1}_{f} | {\cal M}_\mu^{E1,\textrm{IV}} | \Psi^{JM-;0}_i \ra 
= \la \Psi^{1M'+;1}_f | {\cal M}_\mu^{E1,\textrm{IV}} | [\phi^{011+}_d \otimes \Phi^{1-}]^{JM} \ra 
\eeqn{3.31}
where $J$ can be equal to 0, 1 and 2.
Matrix element \rref{2.12} vanishes, 
\beq
\la \Psi^{1M'+;0}_f | {\cal M}_\mu^{E1,\textrm{IV}} | \Psi^{JM-;1}_i \ra = 0.
\eeqn{3.32}
Matrix element \rref{2.13} reads 
\beq
\la \Psi^{1M'+;0}_{f} | {\cal M}_\mu^{E1,\textrm{IS}} | \Psi^{JM-;0}_i \ra 
= \la \Psi^{1M'+;0}_f | {\cal M}_\mu^{E1,\textrm{IS}} | [\phi^{011+}_d \otimes \Phi^{1-}]^{JM} \ra. 
\eeqn{3.33}

When comparing with the microscopic expressions, one observes that important components 
are missing in the $\alpha + n + p$ model. 
The last term of \Eq{2.31} suggests that the transition matrix elements 
involving a virtual excitation of the  $\alpha$ particle described by 
\beq
\la \phi^{1-;1}_{\alpha^* n} | {\cal M}_{\alpha,\mu}^{E1,\textrm{IV}} | \phi^{00+;0}_\alpha \ra
\eeqn{3.40}
could play a significant role. 
Indeed, such a matrix element is related to the giant dipole resonance of the $\alpha$ particle. 
This effect occurs for an initial relative orbital momentum $L = 1$. 

Simulating the effect of matrix element \rref{3.40} is not possible in the present three-body model. 
Indeed, while the value of matrix element \rref{3.40} might be estimated, 
the radial component $g^{1+}_{\alpha^* In}(R)$ of the relative wave function in \Eq{2.31} is unknown. 
\section{Numerical results}
\label{sec:nr}
\subsection{Conditions of the calculations}
The determination of the final $^6$Li$(1^+)$ ground-state wave function 
in a variational calculation is explained in \re{DDB03}. 
The central Minnesota $NN$ potential is employed as neutron-proton interaction \cite{thom77}. 
For the $\alpha+N$ nuclear interaction, the potentials of Voronchev {\it et al} \cite{vor95} 
and of Kanada {\it et al} \cite{KKN79} are employed. 
They are slightly renormalized by respective scaling factors 1.014 and 1.008 
to reproduce the experimental binding energy 3.70 MeV of $^6$Li 
with respect to the $\alpha+n+p$ threshold. 
The Coulomb interaction between $\alpha$ and proton is taken as 
$2e^2\, \mathrm{erf}(0.83\, R)/R$ \cite{RT70}. 
The coupled hyperradial equations are solved with the Lagrange-mesh method \cite{baye15,DDB03}. 
The hypermomentum expansion includes terms up to $K_{\rm max} = 24$, 
which ensures a good convergence of the energy and of the $T = 1$ component of $^6$Li. 
The ground state is essentially $S = 1$ (96 \%). 
The matter r.m.s. radius of the ground state (with 1.4 fm as $\alpha$ radius) 
is found as $\sqrt{r^2} \approx 2.25$ fm with the potential of \re{vor95} 
or 2.24 fm with the potential of \re{KKN79}, \textit{i.e.}\ 
values slightly lower than the experimental value $2.32 \pm 0.03$ fm \cite{exp162}. 
The isotriplet component in the $^6$Li ground state has a squared norm $5.3 \times 10^{-3}$ 
with the potential of \re{vor95} and $4.2 \times 10^{-3}$ with the potential of \re{KKN79}. 

For the initial scattering waves, the radial wave function $u^{011}(r)$ of the deuteron is 
the ground-state solution of the Schr\"odinger equation 
with the Minnesota potential with $\hbar^2/2 m_N=20.7343$ MeV fm$^2$. 
The Schr\"odinger equation is solved by using the Lagrange-Laguerre mesh method \cite{baye15}. 
The converged deuteron energy is $E_d = -2.202$ MeV with 40 mesh points 
and a scaling parameter $h_d = 0.40$. 
The scattering wave functions $g^{L\pi}_i(R)$ of the $\alpha + d$ relative motion 
are calculated with the deep potential of \re{tur15} adapted from the potential of \re{dub94}. 
\subsection{Astrophysical S-factors}
The astrophysical $S$ factor for multipolarity $E\lambda$ is defined 
in terms of the cross section $\sigma_{E\lambda}(E)$ as \cite{Fowler} 
\beq
S_{E\lambda}(E) = E\, \sigma_{E\lambda}(E) \exp(2 \pi \eta),
\eeqn{3.45}
where  $\eta$ is the Sommerfeld parameter. 

\begin{table}
\begin{tabular}{ccc}
\hline
$E$ (MeV) & $S_{E1}^{\textrm{IV}}$ (MeV b) & $S_{E1}^{\textrm{IV+IS}}$ (MeV b)\\
\hline
0.01 & $6.38 \times 10^{-10}$ &  $6.23 \times 10^{-10}$ \\
 0.1 & $1.17 \times 10^{-9}$  &  $1.15 \times 10^{-9}$  \\
   1 & $1.45 \times 10^{-8}$  &  $1.41 \times 10^{-8}$  \\
\hline
\end{tabular}
\caption{$E1$ astrophysical $S$ factor with the isovector (IV) and isovector + isoscalar (IV+IS) models. 
The $\alpha + N$ interaction of \re{vor95} is used.}
\label{tab:1} 
\end{table}
First, we evaluate the role of the two contributions to $S_{E1}$ 
that are calculable in the present model, \textit{i.e.}\ 
the isovector transition involving operator \rref{A5} from the $L = 1$ initial partial wave 
to the $T_f = 1$ component of the $^6$Li ground state 
and the isoscalar transition involving operator \rref{A6} to the $T_f = 0$ component. 
These two contributions add coherently. 
The transition operator given by the first term of \Eq{A6} differs from the ones studied in 
several earlier works \cite{DB86,BZZ90,Ja93}. 
Indeed, it is argued in \re{Ba12} that a neglected term in the matrix element may be rather large in these works. 
In the isoscalar operator \rref{2.3} based on a Siegert transformation 
from which expression \rref{A6} is deduced, 
the second term should be negligible in the present case. 
The resulting difference is that the coefficient of the first term of \Eq{2.3} is smaller by a factor 4 
than in the operators considered in Refs.~\cite{DB86,BZZ90,Ja93}. 

In Table \ref{tab:1}, the resulting isovector and isoscalar $S_{E1}^{\textrm{IV+IS}}$ factor 
is compared at three energies with the purely isovector $S_{E1}^{\textrm{IV}}$ factor. 
The isoscalar correction represents about 2 \%. 
It can be neglected as long as the isovector part is not better known. 
Notice that the isoscalar correction should be more important in the $d(d,\gamma)^4$He capture reaction 
since the photon wavenumber $k_\gamma$ is much larger at low scattering energy. 

\begin{figure}[t]
\setlength{\unitlength}{1 mm}
\begin{picture}(160,105) (0,5) 
\put(0,0){\includegraphics[width=1.0\textwidth]{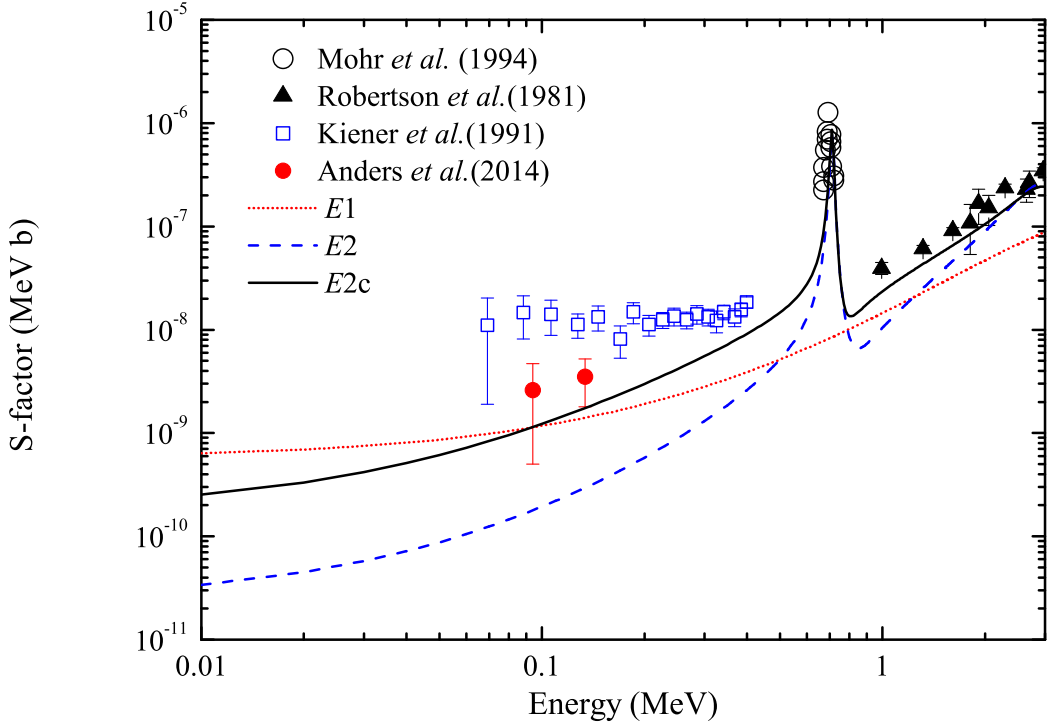}}
\end{picture} \\
\caption{Present $E1$ $S$ factor, $E2$ $S$ factor of \re{TKT16} and corrected $E2$ $S$ factor 
calculated with the $\alpha + N$ potential of \re{vor95} (Model A). 
The experimental data are from Refs.~\cite{robe81} (triangles), \cite{kien91} (squares), 
\cite{mohr94} (open circles), and \cite{luna14} (full circles).}
\label{fig:1}
\end{figure}
With the $\alpha + N$ potential of \re{vor95}, the present IV+IS $S_{E1}$ is represented in Fig.~\ref{fig:1} 
as a dotted line. 
We have reanalyzed $S_{E2}$ calculated with the $E2$ operator of \Eq{A7} within the three-body model of \re{TKT16}, 
depicted as a dashed line in Fig.~\ref{fig:1}. 
At low energies, the cross section is very sensitive to the asymptotic behavior of the overlap integrals 
between the deuteron and the three-body wave functions for partial waves $L = 0$ and 2,
\beq
I_L(R) = \la [\phi^{011} \otimes Y_L(\Omega_R)]^{1M} | \Psi_f^{1M+} \ra, 
\eeqn{3.50}
up to large $\alpha-d$ distances $R$. 
In the model of \re{TKT16}, $I_L(R)$ follows over the interval $[5,10]$ fm 
the expected asymptotic behavior $C_L W_{-\eta_b,L+1/2}(2k_bR)/R$, 
where $\eta_b$ and $k_b$ are the Sommerfeld parameter and wavenumber calculated at the 
separation energy 1.474 MeV of the $^6$Li bound state into $\alpha$ and $d$. 
The $L = 0$ asymptotic normalization coefficient (ANC) is $C_0 \approx 2.05$ fm$^{-1/2}$ in reasonable agreement 
with the value $C_0 \approx 2.30$ fm$^{-1/2}$ extracted in \re{blok93} from experimental data on $\alpha+d$ scattering. 
However, beyond about 10 fm, the absolute value of $I_L(R)$ decreases faster than the correct asymptotics. 
Hence, within that model, $S_{E2}$ is underestimated at low collision energies. 
To solve this problem, beyond $R_0 = 7.5$ fm, we replace $I_L(R)$ 
by the exact asymptotic expression with $C_L$ calculated at 7.5 fm. 
This corrected $S$ factor is denoted as $S_{E2c}$ and is represented as a full line in Fig.~\ref{fig:1}. 
It is significantly larger than $S_{E2}$ because the cross section is sensitive to $R$ values 
up to about 50 fm at $E = 0.1$ MeV. 
From now on, we only use $S_{E2c}$.
Around the resonance, the $S$ factor is dominated by $E2$ transitions. 
Dipole transitions should be dominant below about 0.1 MeV. 

\begin{figure}[thb]
\setlength{\unitlength}{1 mm}
\begin{picture}(160,105) (0,5) 
\put(0,0){\includegraphics[width=1.0\textwidth]{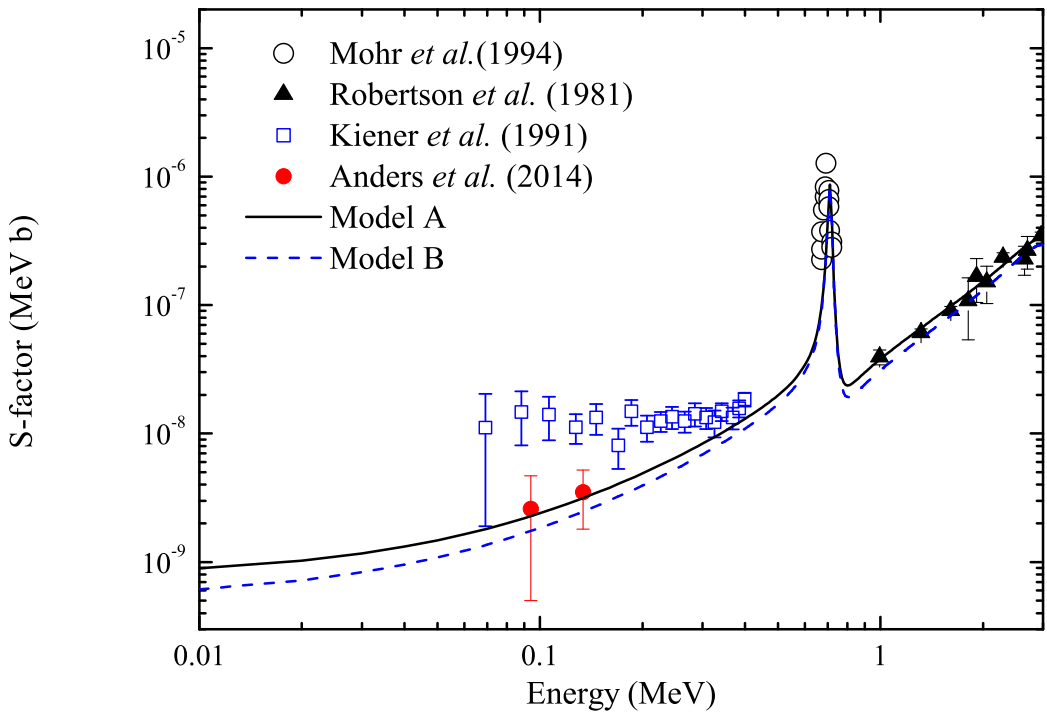}}
\end{picture} \\
\caption{Total $E1+E2c$ astrophysical $S$ factor within the present three-body models A and B. 
The experimental data are from Refs.~\cite{robe81} (triangles), \cite{kien91} (squares), 
\cite{mohr94} (open circles), and \cite{luna14} (full circles).}
\label{fig:2}
\end{figure}

The total $S$ factors $S_{E1}+S_{E2c}$ calculated with the potentials of \re{vor95} (Model A) 
and \re{KKN79} (Model B) are presented in Fig.~\ref{fig:2}. 
They are compared with the direct data of \re{robe81} above the resonance (triangles), 
of \re{mohr94} on resonance (open circles), and of \re{luna14} around 0.1 MeV (full circles). 
The indirect breakup data of \re{kien91} are indicated as squares. 
At low energies, the total $S$ factor obtained in Model A (full line) nicely agrees with the LUNA data. 
The total $S$ factor in Model B (dotted line) is lower by about 35 \% than in Model A (full line) 
but remains within the experimental error bars. 
This relative smallness is related with a smaller $T_f = 1$ component in Model B. 

Despite that several possibly important $T = 1$ contributions are not included in the present discussion, 
\textit{i.e.} mainly the whole $T_i = 1$ component in the initial wave and 
the $T_f = 1$ dipole excitation of the $\alpha$ core in the final wave function, 
one may nevertheless conjecture that a consistent treatment of all isovector $E1$ transitions 
can explain the low-energy experimental data. 
This assumes, however, that the different contributions do not interfere destructively. 
\section{Discussion}
\label{sec:emp}
\subsection{Inadequacy of the exact-masses prescription}
The developments of the previous sections now allow us to discuss 
the validity of the exact-masses prescription. 
We have seen that one can conjecture that isovector $E1$ transitions are able 
to explain the low-energy $S$ factor with a good accuracy. 
This is incompatible with the exact-masses prescription as we now show. 

To simplify the discussion, let us consider $E1$ transitions in the two-body case. 
In the exact-masses prescription, the dimensionless factor $Z_1/A_1 - Z_2/A_2$ 
which multiplies $(A_1 A_2/A) e \cY_{1\mu}(\ve{R})$ in the $E1$ radial operator is replaced by  
\beq
m_N \left(\frac{Z_1}{M_1} - \frac{Z_2}{M_2} \right), 
\eeqn{3.62}
where $M_1$ and $M_2$ are the experimental masses of the colliding nuclei 
and $m_N = \dem(m_n + m_p)$ is the nucleon mass. 
For $N = Z$ nuclei, this factor does not vanish any more in general. 
Notice however that it still vanishes in collisions between identical nuclei. 
It would for example be ineffective to try to describe the forbidden $E1$ deuteron-deuteron capture. 

The factor \rref{3.62} is usually justified by the fact 
that the dipole moment of the nucleus does not vanish in the two-cluster picture with realistic masses. 
It is also sometimes justified by a relativistic correction \cite{noll01}. 
If one replaces the center-of-mass coordinates of the clusters by center-of-energy coordinates, 
the electric dipole moment becomes closer to expression \rref{3.62}. 
Though it is true that relativistic corrections could play a role, 
the argument is weakened by the fact that the original factor $Z_1/A_1 - Z_2/A_2$ 
is based on a microscopic description in terms of nucleons 
while the center-of energy argument is based on a two-cluster structure. 
Consistent relativistic corrections should also be based on nucleons. 

The mass of a nucleus $^A_Z$X$_N$ can be written as  
\beq
M = Am_N + (N-Z) \dem (m_n - m_p) - B(A,Z)/c^2,
\eeqn{3.61}
where $B(A,Z)$ is the binding energy. 
As the binding energy per nucleon is small with respect to the nucleon mass energy, 
factor \rref{3.62} can be approximated for a capture involving nuclei with $N = Z$ as 
\beq
m_N \left(\frac{Z_1}{M_1} - \frac{Z_2}{M_2} \right) 
& \approx & \frac{Z_1}{A_1} \left(1 + \frac{B(A_1,Z_1)}{A_1 m_N c^2} \right) 
- \frac{Z_2}{A_2} \left(1 + \frac{B(A_2,Z_2)}{A_ 2 m_N c^2} \right) 
\eol
& = & \frac{1}{2m_N c^2} \left(\frac{B(A_1,Z_1)}{A_1} - \frac{B(A_2,Z_2)}{A_ 2} \right).
\eeqn{3.63}
This correction is small since the binding energy per nucleon does not vary much 
from one nucleus to another. 
In the $\alpha+d$ case, it is about $4 \times 10^{-4}$. 
This factor is quite small and is fortuitously able to reproduce 
a plausible order of magnitude of forbidden $E1$ transitions. 
However, there is no physical relation between this correction and 
the dominant isovector transitions when the $E1$ transition is isospin forbidden. 
Indeed, the binding energy per nucleon of a $N = Z$ nucleus mainly depends 
on the dominant $T = 0$ component of its ground state. 
It is in no appreciable way sensitive to $T = 1$ admixtures 
as $E1$ matrix elements describing an isospin-forbidden capture should be. 

Can the exact-masses prescription give a realistic energy dependence of the $S$ factor 
below the 711 keV resonance? 
Since the dominant initial orbital momentum is $l = 1$, 
the low-energy dependence of the initial relative scattering wave [\Eq{2.20}] is 
close to the dependence of the regular Coulomb function $F_1$ (see Eq.~(7) of \re{bb00}), 
\beq
g_i^{J-}(R) \approx E^{1/4} \left[f_0(R) + f_1(R) E + \dots\right] \exp(-\pi\eta).
\eeqn{3.42}
In any model, the coefficients $f_i(R)$ are calculable functions of $R$. 
For Coulomb waves, they are given by Eq.~(22) of \re{bb00}. 
The integral $M(E)$ over $R$ appearing in matrix element \rref{3.31} and its various corrections 
can thus be written at very low energies as 
\beq
M(E) \propto E^{1/4} \left(M_0 + M_1 E + \dots\right) \exp(-\pi\eta),
\eeqn{3.44}
where coefficient $M_i$ is an integral involving $f_i(R)$, the radial operator $R$, 
and the overlap integral $I_L(R)$ of the bound-state wave function 
with the internal cluster wave functions (such as \Eq{3.50} in the three-body case). 
This last factor is quite different in the exact-masses prescription 
(where it is just given by the final bound-state wave function with $T_f = 0$) 
and in isovector matrix elements (where it corresponds to a small $T_f = 1$ admixture 
of the final wave function). 
In particular, it is quite different at large distances since the $T_f = 1$ admixture 
does not have an $\alpha+d$ asymptotic behavior. 
Hence $M_0$ and $M_1$ may be quite different in both descriptions. 

The low-energy behavior of the $S$ factor is given by the expansion 
\beq
S(E) = S(0) (1 + s_1 E + \dots),
\eeqn{3.46}
where the slope $s_1$ depends on the ratio of $M_1$ and $M_0$ \cite{bb00,Ba13}.  
At sufficiently low energies, this ratio computed with the exact-masses prescription 
is not related to the one in the isovector-transition picture. 
The prescription is not expected to reproduce the physical energy slope of $S_{E1}$ near zero energy. 
\subsection{Role of $S$-wave capture}
\label{sec:S}
The $E1$ $S$ factor which is dominant below about 0.1 MeV decreases with decreasing energy 
since it is due to a transition from an initial $P$ wave. 
As transitions from $S$ waves have an almost flat energy dependence at very low energies, 
an energy (possibly very low) must exist where transitions from an initial $S$ wave dominate. 

The $E2$ capture cross section mainly corresponds to a transition between 
an initial $D$ wave and the $^6$Li ground state. 
In the present $\alpha + n + p$ model, an $E2$ capture from an initial $S$ wave exists 
but is smaller than the other $E2$ contributions by several orders of magnitude 
in the energy range of Figs.~\ref{fig:1} and \ref{fig:2} \cite{TKT16}. 
However, other transitions starting from the $S$ wave are possible, 
which are not considered here. 
Since the $^6$Li, $^4$He, and $^2$H ground states contain a $D$-wave component 
due to the $NN$ tensor force, 
several types of $E2$ transition from an initial $S$ wave can contribute. 
As the energy dependence of transition matrix elements from an initial $S$ wave 
is much weaker than for a $D$ wave, this contribution should become dominant 
below some low energy. 
This mechanism is well illustrated by the $d(d,\gamma)^4$He capture reaction \cite{AAS11,DBS14}. 
The main contribution to the capture at low energies is due to the small $D$-wave 
components of the $\alpha$ particle and of the deuterons. 
For $^4$He$(d,\gamma)^6$Li, earlier works indicate that this component is small \cite{LR86,CET90} 
but they are restricted to energies above the 711 keV resonance. 
It is thus not possible for the moment to estimate the energy below which this mechanism 
would be important nor the order of magnitude of its contribution to the cross section at low energies. 

We have performed a partial test within the $\alpha + n + p$ three-body model 
by including a $D$-wave component in the initial deuteron wave function. 
With the full deuteron wave function obtained with the soft-core potential of \re{Re68}, 
the $S$-wave contribution to $S_{E2}$ is negligible above 10 keV. 
The resulting $S$-wave capture remains very small in agreement with previous studies. 
Full confirmation requires a calculation taking simultaneous account 
of the $^6$Li, $^4$He, and $^2$H $D$ components. 
Such a calculation requires extensions of the three-body model 
but is within the reach of present-day {\it ab initio} calculations. 

The magnetic dipole capture is another case where capture from the $S$ wave can occur. 
The microscopic $M1$ operator can be written as a sum of a term proportional 
to the total angular momentum and a residual spin term. 
The matrix elements of the first term must vanish in any model because of the orthogonality 
between the initial and final wave functions \cite{DB83,TBL91,noll01}. 
It is thus meaningless to evaluate $M1$ capture in models (like the present one) 
where the initial scattering partial waves and the final bound-state wave function 
are not derived from the same Hamiltonian. 
When the matrix element of the residual spin term is small, 
$M1$ transitions are strongly hindered. 
The energy below which $M1$ transitions dominate $E1$ transitions must be very small.  
\section{Conclusion}
\label{sec:conc}
In this paper, we discuss the properties expected for a realistic treatment 
of the isospin-forbidden $E1$ component of the $\alpha(d,\gamma)^6$Li reaction. 
Since such a calculation is presently not available at the nucleon microscopic level, 
we evaluate some contributions that are accessible with a three-body model. 
The higher-order contribution from the isoscalar part of the operator is found small 
and could be neglected in future calculations of this reaction to a good approximation. 
The isotriplet component of the final $^6$Li$(1^+)$ ground state due to deuteron virtual excitations 
leads to a total $E1+E2$ astrophysical $S$ factor compatible with the experimental data 
at low energies of \re{luna14}. 
Other $E1$ components of the $S$ factor due to similar distortions of the initial scattering wave and 
to $T_\alpha = 1$ virtual excitations of the $\alpha$ particle in the $^6$Li ground state 
are not accessible within the present model. 
We conjecture that, with these other contributions, 
isovector transitions are able to explain the data without adjustable parameter. 
We also emphasize the need for correct $\alpha + d$ asymptotics of the three-body wave function 
to correctly describe the $E2$ component of the astrophysical $S$ factor.

We have questioned the exact-masses prescription of the potential model 
and shown that it is not founded at the microscopic level. 
It is incompatible with an explanation of the low-energy data 
in terms of isovector $E1$ transitions. 
Its order of magnitude and energy dependence may be accidentally correct 
but this prescription does not seem to have a physical meaning. 
Its use should be avoided in capture reactions between $N = Z$ nuclei 
such as $\alpha(d,\gamma)^6$Li or $^{12}$C($\alpha,\gamma)^{16}$O. 

Radiative capture from the $S$ wave should become dominant below some unknown low energy. 
It is not completely established that this type of transition is too weak 
to contribute to the capture process at the lowest energies where experiments are available. 
This initial partial wave can play a role in $M1$ and $E2$ transitions. 
While $M1$ transitions are strongly hindered by the orthogonality between the initial and final states, 
it could be worth reexamining the $E2$ radiative capture at very low energies 
to evaluate the role of the various $D$-wave components in the initial and final clusters. 
Indeed such components in $^2$H, $^4$He, and $^6$Li render possible transitions 
from an initial $S$ wave with a much weaker energy dependence at very low energies 
as obtained in the $d(d,\gamma)^4$He reaction \cite{AAS11}. 

As long as {\it ab initio} calculations or advanced microscopic cluster calculations 
involving various forms of isospin mixing are not available, 
the importance of $E1$ transitions in the $\alpha +d \rightarrow\, ^6$Li$ + \gamma$ reaction 
will remain poorly known. 
The three-body model is interesting as it offers simpler physical interpretations 
than more elaborate models. 
Some aspects of the present three-body study, however, limit its predictive power. 
Extensions are possible which should be considered in the future. 
The first one is to improve the $\alpha+d$ asymptotics of the final $^6$Li wave function. 
A second one is to replace the frozen-deuteron description in the initial wave 
by a flexible three-body description allowing distortions of the deuteron and, 
in particular, the appearance of isotriplet admixtures which will contribute to $E1$ capture 
in a consistent way with those of the final $^6$Li ground state. 
A third, more difficult, extension would involve core excitations, 
\textit{i.e.}, additional configurations for the $\alpha$ particle. 
We expect that a significant component of $E1$ capture could come from $T = 1$ 
virtual excitations of the $\alpha$ particle corresponding to its giant dipole resonance. 
Future three-body but also microscopic calculations of $E1$ $\alpha + d$ capture 
should usefully include this kind of configuration. 
\acknowledgments
E.M.T thanks the Fonds de la Recherche Scientifique - FNRS (Belgium) for a grant. 
He is grateful to P. Descouvemont for his kind invitation and welcome. 
He also acknowledges useful discussions with L.D.~Blokhintsev and A.S.~Kadyrov. 

\end{document}